\documentclass[Physsubmission, Phys]{SciPost}
\hypersetup{
    colorlinks,
    linkcolor={red!50!black},
    citecolor={blue!50!black},
    urlcolor={blue!80!black}
}

\usepackage[bitstream-charter]{mathdesign}
\DeclareSymbolFont{usualmathcal}{OMS}{cmsy}{m}{n}
\DeclareSymbolFontAlphabet{\mathcal}{usualmathcal}

\begin{document}
\begin{center}{\Large \textbf{
Measurement of Intermittency for Charged Particles in Au + Au Collisions at $\sqrt{s_\mathrm{NN}}$ = 7.7-200 GeV from STAR\\
}}\end{center}

\begin{center}
Jin Wu\textsuperscript{1$\star$} (for the STAR Collaboration)
\end{center}

% TODO: write all affiliations here.
% Format: institute, city, country
\begin{center}
{\bf 1} Central China Normal University \\
% TODO: provide email address of corresponding author
* wuj276@mails.ccnu.edu.cn
\end{center}

\begin{center}
\today
\end{center}

% For convenience during refereeing (optional),
% you can turn on line numbers by uncommenting the next line:
%\linenumbers
% You should run LaTeX twice in order for the line numbers to appear.

\definecolor{palegray}{gray}{0.95}
\begin{center}
\colorbox{palegray}{
  \begin{tabular}{rr}
  \begin{minipage}{0.1\textwidth}
    \includegraphics[width=30mm]{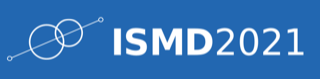}
  \end{minipage}
  &
  \begin{minipage}{0.75\textwidth}
    \begin{center}
    {\it 50th International Symposium on Multiparticle Dynamics}\\ {\it (ISMD2021)}\\
    {\it 12-16 July 2021} \\
    \doi{10.21468/SciPostPhysProc.?}\\
    \end{center}
  \end{minipage}
\end{tabular}
}
\end{center}

%\linenumbers
\section*{Abstract}
{\bf
Local density fluctuations near the QCD critical point can be probed by intermittency analysis of scaled factorial moments in relativistic heavy-ion collisions. We report the first measurement of intermittency for charged particles in Au + Au collisions at $\sqrt{s_\mathrm{NN}}$ = 7.7-200 GeV from the STAR experiment at RHIC. We observe scaling behaviors in central Au + Au collisions, with the extracted scaling exponent decreasing from mid-central to the most central Au + Au collisions. Furthermore, the scaling exponent exhibits a non-monotonic energy dependence with a minimum around $\sqrt{s_\mathrm{NN}}$ = 20-30 GeV in central Au + Au collisions.
}

% TODO: include a table of contents (optional)
% Guideline: if your paper is longer that 6 pages, include a TOC
% To remove the TOC, simply cut the following block
\vspace{10pt}
\noindent\rule{\textwidth}{1pt}
\tableofcontents\thispagestyle{fancy}
\noindent\rule{\textwidth}{1pt}
\vspace{10pt}

\section{Introduction}
\label{sec:intro}
The major goal of the Beam Energy Scan (BES) at the Relativistic Heavy Ion Collider (RHIC) is to explore the phase diagram of quantum chromodynamics (QCD)~\cite{QCDReport,STARPRLMoment}. An important landmark of the QCD phase structure is the critical point (CP), which is the end point of first-order phase boundary between quark-gluon and hadronic phases~\cite{CEP2}. In the thermodynamic limit, the correlation length diverges at the CP and the system becomes scale invariant and fractal~\cite{invariant3}. It is shown that the density fluctuations near the QCD critical point form a distinct pattern of power-law or intermittent behavior in the matter produced in high energy heavy-ion collisions~\cite{AntoniouPRL}.

In analogy to the critical opalescence observed in conventional matter near the critical point, the related fractal and self-similar geometry of QCD matter will lead to local density fluctuations that obey intermittent behavior~\cite{AntoniouPRL}. Based on the effective action belonging to three-dimensional Ising universality class, the intermittency of QCD matter is revealed in transverse momentum spectra as a power-law (scaling) behavior of scaled factorial moment (SFM) in heavy-ion collisions~\cite{AntoniouPRL}. An intermittent behavior has observed in Si + Si collisions at 158A GeV from the NA49 experiment~\cite{NA49EPJC}. Meanwhile, studies based on a critical Monte Carlo with self-similar property~\cite{CMCPLB} and transport model with hadronic potentials~\cite{UrQMDLi} demonstrate that the intermittency could be visible in Au + Au collisions at RHIC energies.

\section{Analysis Details}
In high-energy experiments, local power-law fluctuations can be detectable through the measurements of scaled factorial moment (SFM) which is defined as:\\
\begin{equation}
F_{q}(M)=\frac{\langle\frac{1}{M^{D}}\sum_{i=1}^{M^{D}}n_{i}(n_{i}-1)\cdots(n_{i}-q+1)\rangle}{\langle\frac{1}{M^{D}}\sum_{i=1}^{M^{D}}n_{i}\rangle^{q}},
 \label{Eq:FM}
\end{equation}

\noindent where $M^{D}$ is the number of cells in D-dimensional momentum space, $n_{i}$ is the measured multiplicity in the $i$-th cell, and $q$ is the order of moment.

Another expected power-law behavior that describes relationship between $F_{q}(M)$ and $F_{2}(M)$ is defined as~\cite{GLPRL,GLPRD}:
\begin{equation}
F_{q}(M)\propto F_{2}(M)^{\beta q}.
 \label{Eq:FqF2scaing}
\end{equation}

Moreover, the scaling exponent$\nu$ quantitatively describes the values of $\beta_{q}$:
\begin{equation}
\beta_{q} \propto (q-1)^{\nu}.
 \label{Eq:betaqscaing}
\end{equation}

Here $\nu$ specifies scaling (power-law) behavior of $F_{q}(M)$. According to Ginzburg-Landau (GL) theory, the critical $\nu$ is equal to 1.304 in entire space phase~\cite{GLPRL}, while it is equal to 1.0 from the two-dimensional Ising model~\cite{GLPRD}. 

\begin{figure*}[htp]
     \centering
     \includegraphics[scale=0.65]{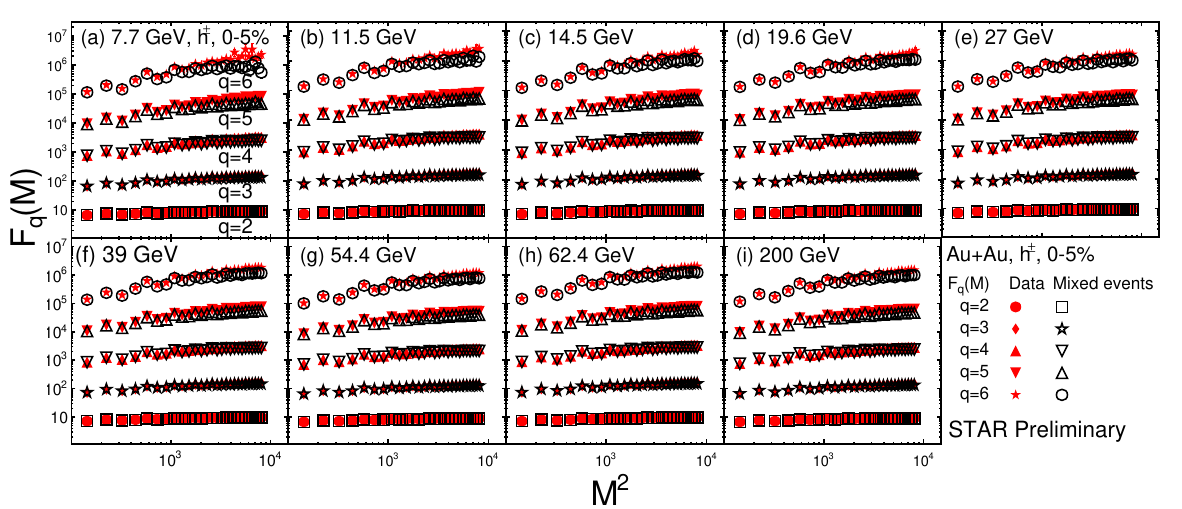}
     \label{Fig:SFM}
     \caption{ $F_{q}(M)$ (up to sixth order) of charged particles in transverse momentum space for the most central (0-5\%) Au + Au collisions at $\sqrt{s_\mathrm{NN}}$ = 7.7-200 GeV in double-logarithmic scale.}
\end{figure*}

The data reported here were obtained from Au + Au collisions at $\sqrt{s_\mathrm{NN}}$ = 7.7, 11.5, 14.5, 19.6, 27, 39, 54.4, 62.4 and 200 GeV, which were recorded by the STAR experiment at RHIC from 2010 to 2017. Protons ($p$), antiprotons ($\bar{p}$), kaons ($K^{\pm}$) and pions ($\pi^{\pm}$) are analyzed as charged particles, and their identifications are carried out using the Time Projection Chamber (TPC) and the Time-of-Flight (TOF) detectors. To avoid the self-correlation, the centrality was determined from uncorrected charged particles within a pseudo-rapidity window of $0.5<\mid \eta \mid <1$, which was chosen to be beyond the analysis window of $\mid \eta \mid <0.5$.

To subtract the background at the level of SFM, a correlator $\Delta F_{q}(M)$ is defined in terms of original and mixed events, i.e., $\Delta F_{q}(M)=F_{q}(M)^{data}-F_{q}(M)^{mix}$~\cite{NA49EPJC}. In addition, a cell-by-cell method is proposed for efficiency correction on SFM~\cite{RefEfficiency}. The statistical uncertainties are estimated by Bootstrap method, and the systematic uncertainties are estimated by varying the experimental requirements for tracks in the TPC and TOF.

\section{Results and Discussion}
\label{sec:another}

Figure 1 shows $F_{q}(M)^{data}$ and $F_{q}(M)^{mix}$, from the second order to the sixth order in the most central (0-5\%) collisions for various $\sqrt{s_\mathrm{NN}}$. Based on the statistics of BES-I data, $F_{q}(M)$ can be calculated in the range of $M^{2}$ from 1 to $100^{2}$ and up to the sixth order (q=6). It is observed that $F_{q}(M)^{data}$ is larger than $F_{q}(M)^{mix}$ at large $M^{2}$ region for various $\sqrt{s_\mathrm{NN}}$, thus a deviation of $\Delta F_{q}(M)$ from zero is present in central Au + Au collisions.

%%%%%%%%%%%%%%%Fig2 Fq(M)/F2(M)
\begin{figure*}[!htp]
     \centering
     \includegraphics[scale=0.65]{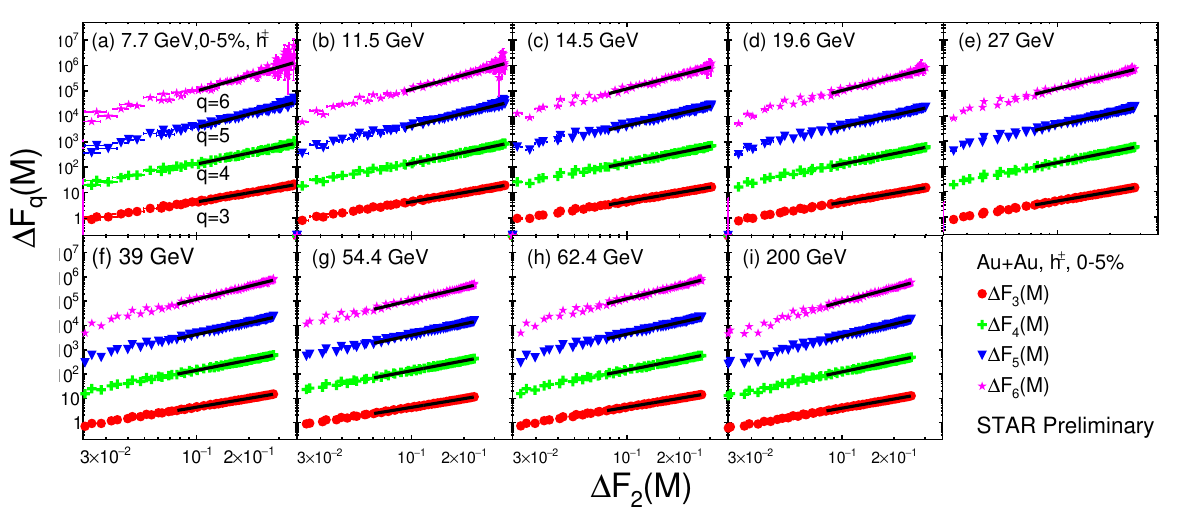}
     \label{Fig:FqF2}
     \caption{$\Delta F_{q}(M)$ (q=3-6) as a function of $\Delta F_{2}(M)$ in the most central (0-5\%) Au + Au collisions at $\sqrt{s_\mathrm{NN}}$ = 7.7-200 GeV in double-logarithmic scale.}
\end{figure*}

Figure 2 shows $\Delta F_{q}(M)$ (q=3-6), as a function of $\Delta F_{2}(M)$ in the most central (0-5\%) collisions for various $\sqrt{s_\mathrm{NN}}$. We clearly observe that the correlators $\Delta F_{q}(M)$ (q=3-6) exhibit scaling behavior with $\Delta F_{2}(M)$. 

The value of $\beta_{q}$ is obtained through a power-law fit of Eq.~\eqref{Eq:FqF2scaing} as shown in Figure 2, and its statistical error is determined by the fit. Figure 3(a) shows $\beta_{q}$ as a function of $q-1$ in the most central Au + Au collisions for $\sqrt{s_\mathrm{NN}}$ = 7.7-200 GeV. Consistent with theoretical expectation, $\beta_{q}$ also obeys a good scaling behavior with q, thus $\nu$ can be obtained through a power-law fit of Eq.~\eqref{Eq:betaqscaing}. Figure 3(b) shows the extracted $\nu$ as a function of $\langle N_{part} \rangle$ in central Au + Au collisions at various $\sqrt{s_\mathrm{NN}}$. We find that $\nu$ decreases from mid-central (30-40\%) to the most central (0-5\%) Au + Au collisions.

Figure 4 shows the energy dependence of $\nu$ of charged particles in central Au + Au collisions at $\sqrt{s_\mathrm{NN}}$ = 7.7-200 GeV. It is observed that the $\nu$ exhibits a non-monotonic behavior on collision energy and seems to reach a minimum around $\sqrt{s_\mathrm{NN}}$ = 20-30 GeV. Higher statistics data from BES-II will help to confirm the trend of energy dependence of $\nu$.

\begin{figure*}[!htp]
     \centering
     \includegraphics[scale=0.50]{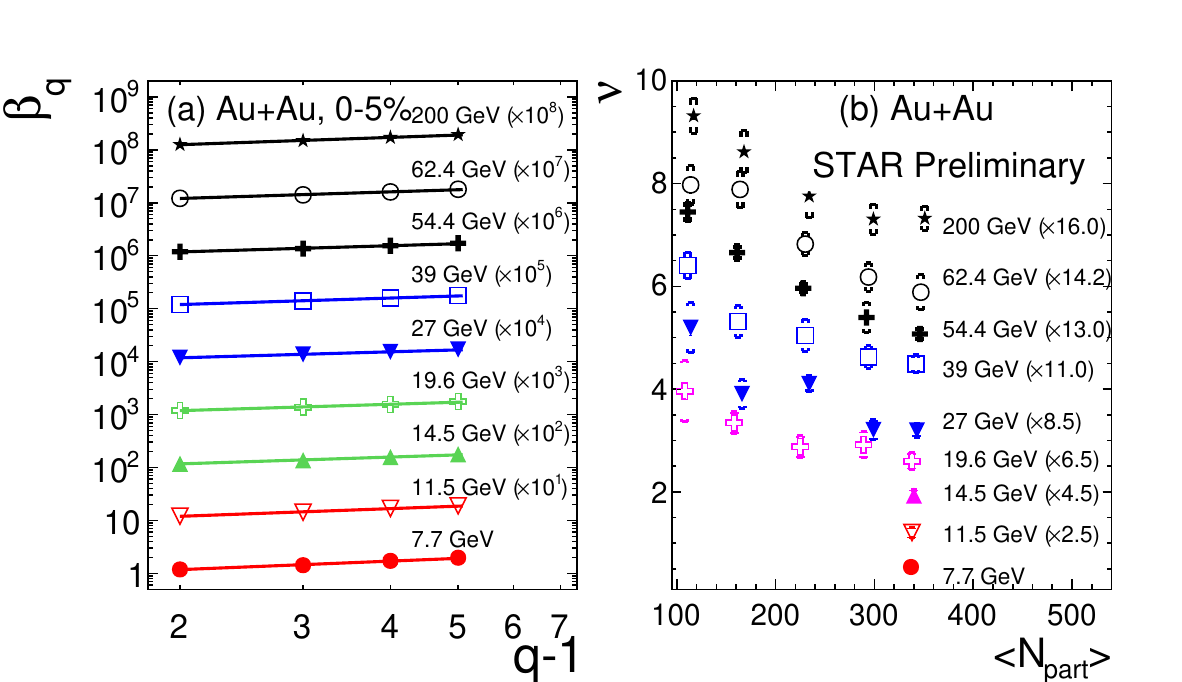}
     \label{Fig:bataq}
     \caption{(a) $\beta_{q}$ (q=3-6) as a function of q-1 in most central Au + Au collisions at $\sqrt{s_\mathrm{NN}}$ = 7.7-200 GeV. (b) $\nu$ as a function of $\langle N_{part}\rangle$ in central Au + Au collisions.}
\end{figure*}

\vspace{-0.5cm}
\begin{figure*}[!htp]
     \centering
     \includegraphics[scale=0.30]{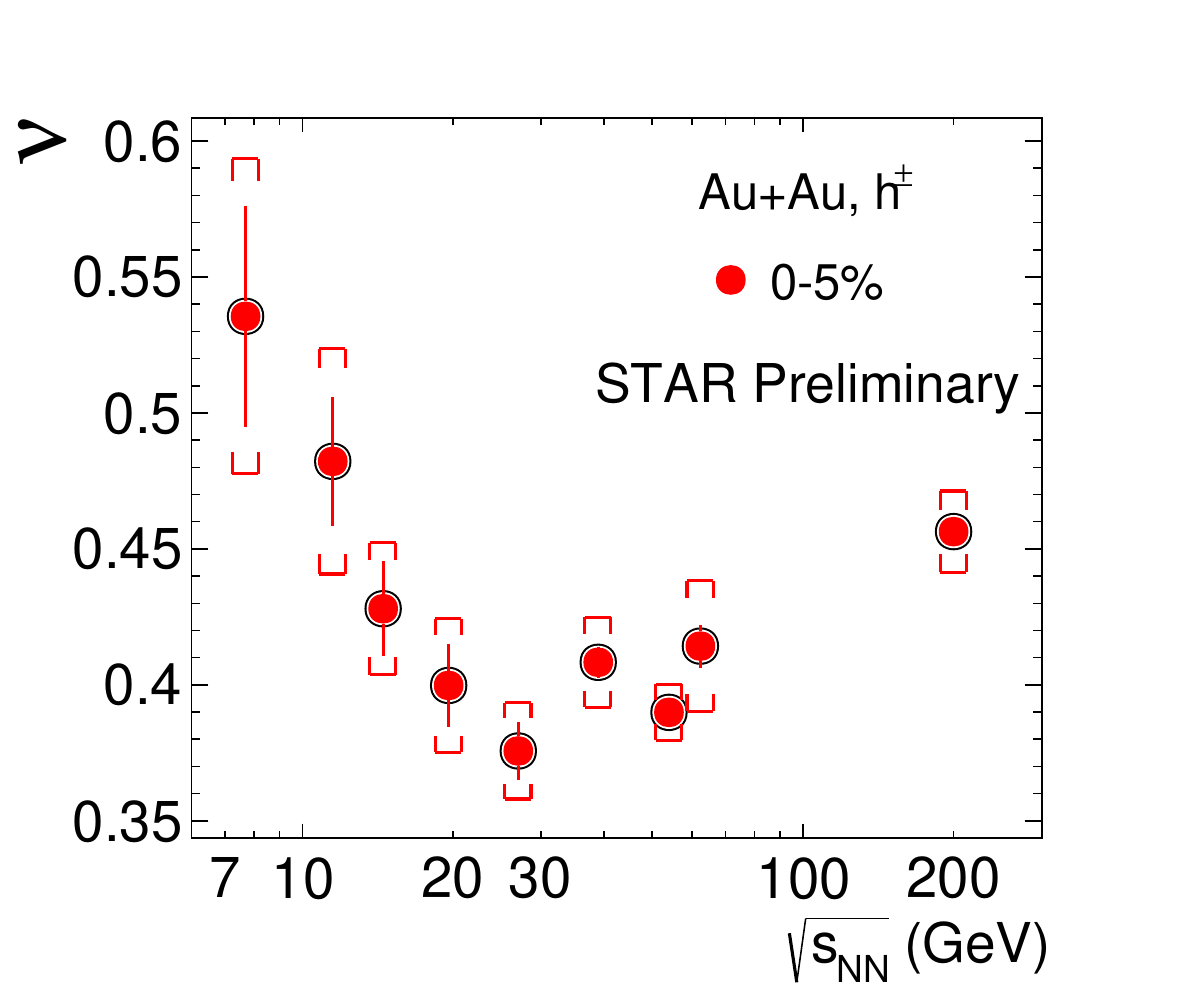}
     \label{Fig:nuenergy}
     \caption{Energy dependence of $\nu$ for charged particles in Au + Au collisions at $\sqrt{s_\mathrm{NN}}$ = 7.7- 200 GeV. The statistical and systematic errors are shown in bars and brackets, respectively.}
\end{figure*}

\vspace{-0.5cm}
\section{Summary}
In summary, we report the first measurements of intermittency for charged particles in Au + Au collisions at $\sqrt{s_\mathrm{NN}}$ = 7.7-200 GeV from the STAR experiment. Scaled factorial moments (up to the sixth order) for $p$, $\bar{p}$, $K^{\pm}$ and $\pi^{\pm}$ within $|\eta|<0.5$, have been measured in available transverse momentum space. Scaling behavior is clearly visible in Au + Au collisions which is consistent with theoretical predictions. The scaling exponent is related to the critical component, and we observe that it shows a non-monotonic behavior on $\sqrt{s_\mathrm{NN}}$ with a dip around 20-30 GeV in the most central (0-5\%) Au + Au collisions. This non-monotonic behavior needs to be understood with more theoretical inputs. With significantly improved statistics, the RHIC BES Phase-II program will allow for a more precise measurement of intermittency in heavy-ion collisions.

\vspace{-0.5cm}
\section*{Acknowledgements}
This work is supported by the National Key Research and Development Program of China (Grants No. 2020YFE0202002), the National Natural Science Foundation of China (Grants No. 12122505). And the Ministry of Science and Technology (MoST) under grant No. 2016YFE0104800 are also acknowledged.
\vspace{-0.5cm}
\bibliography{SciPost_Example_BiBTeX_File.bib}

\begin{thebibliography}{10}
\providecommand{\url}[1]{\texttt{#1}}
\providecommand{\urlprefix}{URL }
\expandafter\ifx\csname urlstyle\endcsname\relax
  \providecommand{\doi}[1]{doi:\discretionary{}{}{}#1}\else
  \providecommand{\doi}{doi:\discretionary{}{}{}\begingroup
  \urlstyle{rm}\Url}\fi
\providecommand{\eprint}[2][]{\url{#2}}

\bibitem{QCDReport}
A.~Bzdak \emph{et~al.},
\newblock \emph{{Mapping the Phases of Quantum Chromodynamics with Beam Energy
  Scan}},
\newblock Phys. Rept. \textbf{853}, 1 (2020),
\newblock \doi{10.1016/j.physrep.2020.01.005}.

\bibitem{STARPRLMoment}
J.~Adam~{\it et al.} (STAR~Collaboration),
\newblock \emph{{Nonmonotonic Energy Dependence of Net-Proton Number
  Fluctuations}},
\newblock Phys. Rev. Lett. \textbf{126}(9), 092301 (2021),
\newblock \doi{10.1103/PhysRevLett.126.092301}.

\bibitem{CEP2}
Y.~Hatta and M.~A. Stephanov,
\newblock \emph{{Proton number fluctuation as a signal of the QCD critical
  endpoint}},
\newblock Phys. Rev. Lett. \textbf{91}, 102003 (2003),
\newblock \doi{10.1103/PhysRevLett.91.102003}.

\bibitem{invariant3}
E.~A. De~Wolf \emph{et~al.},
\newblock \emph{{Scaling laws for density correlations and fluctuations in
  multiparticle dynamics}},
\newblock Phys. Rept. \textbf{270}, 1 (1996),
\newblock \doi{10.1016/0370-1573(95)00069-0}.

\bibitem{AntoniouPRL}
N.~G. Antoniou \emph{et~al.},
\newblock \emph{{Critical opalescence in baryonic QCD matter}},
\newblock Phys. Rev. Lett. \textbf{97}, 032002 (2006),
\newblock \doi{10.1103/PhysRevLett.97.032002}.

\bibitem{NA49EPJC}
T.~Anticic~{\it et al.} (NA49~Collaboration),
\newblock \emph{{Critical fluctuations of the proton density in A+A collisions
  at 158$A$ GeV}},
\newblock Eur. Phys. J. C \textbf{75}(12), 587 (2015),
\newblock \doi{10.1140/epjc/s10052-015-3738-5}.

\bibitem{CMCPLB}
J.~Wu \emph{et~al.},
\newblock \emph{{Probing QCD critical fluctuations from intermittency analysis
  in relativistic heavy-ion collisions}},
\newblock Phys. Lett. B \textbf{801}, 135186 (2020),
\newblock \doi{10.1016/j.physletb.2019.135186}.

\bibitem{UrQMDLi}
P.~Li \emph{et~al.},
\newblock \emph{{Proton correlations and apparent intermittency in the UrQMD
  model with hadronic potentials}},
\newblock Phys. Lett. B \textbf{818}, 136393 (2021),
\newblock \doi{10.1016/j.physletb.2021.136393}.

\bibitem{GLPRL}
R.~C. Hwa and M.~T. Nazirov,
\newblock \emph{{Intermittency in second order phase transition}},
\newblock Phys. Rev. Lett. \textbf{69}, 741 (1992),
\newblock \doi{10.1103/PhysRevLett.69.741}.

\bibitem{GLPRD}
R.~C. Hwa,
\newblock \emph{{Scaling exponent of multiplicity fluctuation in phase
  transition}},
\newblock Phys. Rev. D \textbf{47}, 2773 (1993),
\newblock \doi{10.1103/PhysRevD.47.2773}.

\bibitem{RefEfficiency}
J.~Wu \emph{et~al.},
\newblock \emph{{Intermittency analysis of proton numbers in heavy-ion
  collisions at energies available at the BNL Relativistic Heavy Ion
  Collider}},
\newblock Phys. Rev. C \textbf{104}(3), 034902 (2021),
\newblock \doi{10.1103/PhysRevC.104.034902}.

\end{thebibliography}

\nolinenumbers

\end{document}